\documentclass[aps,amssymb,10pt,showpacs,showkeys,letterpaper]{revtex4}
\usepackage{graphicx}
\usepackage{amsmath}
\usepackage{epsfig}
\usepackage{bm}

\newcommand{\be}{\begin{equation}}
\newcommand{\ee}{\end{equation}}
\newcommand{\beq}{\begin{eqnarray}}
\newcommand{\eeq}{\end{eqnarray}}

\begin{document}

\title{Geodesic Structure of Lifshitz Black Holes in 2+1 Dimensions}

\author{ Norman Cruz }
 \email{norman.cruz@usach.cl}
\affiliation{\it Departamento de F\'{\i}sica, Facultad de Ciencia, Universidad
de Santiago de Chile, Casilla 307,\\
Santiago, Chile}

\author{ Marco Olivares }
\email{marco.olivaresrubilar@gmail.com}
\affiliation{\it Instituto de F\'{\i}sica,\\ Pontificia Universidad de
Cat\'{o}lica de Valpara\'{\i}so, Av. Universidad 330, Curauma,\\
Valpara\'{\i}so, Chile}

\author{ J. R. Villanueva }

 \email{jose.villanuevalob@uv.cl}
\affiliation{ Departamento de F\'{\i}sica y Astronom\'ia, Facultad
de Ciencias, Universidad de Valpara\'iso, Gran Breta\~na 1111,\\
Valpara\'iso, Chile,}
\affiliation{Centro de Astrof\'isica de Valpara\'iso, \\Gran Breta\~na 1111, Playa Ancha,
\\ Valpara\'{\i}so, Chile.}

\date{\today}

\begin{abstract}
We present a study of the geodesic equations of a black hole
space-time which is a solution of the three-dimensional NMG theory
and is asymptotically Lifshitz with $z=3$ and $d=1$ as found in
[Ayon-Beato E., Garbarz A., Giribet G. and Hassaine M., Phys. Rev.
{\bf D} 80, 104029 (2009)]. By means of the corresponding effective
potentials for massive particles and photons we find the allowed
motions by the energy levels. Exact solutions for radial and non-radial 
geodesics are given in terms of the Weierstrass elliptic
$\wp$, $\sigma$, and $\zeta$ functions.
\end{abstract}

\pacs{04.20.Fy, 04.20.Jb, 04.40.Nr, 04.70.Bw}

\keywords{Black Holes;  Elliptic Functions.}

\maketitle

\tableofcontents


\section{Introduction}
The New Massive Gravity (NMG)~\cite{NMG} theory has deserved special
attention due to its remarkable physical properties. Since lower-dimensional 
gravity has been an active field of research in order to
shed some light on the behavior of classical and quantum aspects of
this interaction, models in (2+1) dimensions have attracted a
lot of research. In the case of NMG its (2+1)-dimensional version
is described by an action defined by
\begin{equation}
S=\frac{1}{16\pi G_{3}}\int d^{3}x\sqrt{-g}\left[ R-2\lambda
-\frac{1}{m^{2}} \left( R_{\mu \nu }R^{\mu \nu
}-\frac{3}{8}R^{2}\right) \right] , \label{actionNMG}
\end{equation}
where $G_{3}$ is the three-dimensional Newton constant, $m$ is the
"relative mass parameter, and $\lambda$ is the three-dimensional
cosmological constant.  The minus sign in the Einstein-Hilbert term avoids
negative mass and entropy.  This action gives rise to field
equations with a second order trace.  Its linearized version is
equivalent to the Fierz-Pauli action for a massive spin-2 field in
three dimension, so it is a unitary model~\cite{Nakasone}. This theory
admits black hole-type solutions asymptotically Lifshitz,
characterized by the metrics~\cite{Kachru}
\begin{equation}
ds^2=-\frac{r^{2z}}{\ell^{2z}}dt^2+\frac{\ell^2}{r^2}dr^2
+\frac{r^2}{\ell^2}d\vec{x}^2,\label{ansatz0}
\end{equation}
where $\vec{x}$ is a $d$-dimensional vector. These space-times
present invariance under the rescaling
$(t,\vec{x},r)\mapsto(\lambda^{z}t,\lambda\vec{x},\lambda^{-1}r)$.
Many investigations have addressed to find black hole solutions
which asymptote the metrics given by (\ref{ansatz0}).
A four-dimensional topological black hole which is asymptotically
Lifshitz with dynamical exponent $z=2$ was found in~\cite{Mann}.
Other Lifshitz black hole solutions were investigated
in~\cite{DanielssonI,DanielssonII}. A black hole solution for $z=2$
in $d=2$ was found in~\cite{Balasu}.

In this paper our aim is to investigate the geodesic
structure of black hole space-time which is a solution of the
three-dimensional NMG theory and is asymptotically Lifshitz with
$z=3$ and $d=1$~\cite{AyonBeato:2009nh}. Similar studies have been
made for Topological Lifshitz black hole in $3+1$
dimensions~\cite{germancito}. We investigate all the possible
movements allowed in this space-time by means of a detailed analysis
of the effective potentials. We find the exact solutions describing
the path of massive particles and photons. In order to obtain a
direct visualization of the allowed motions, we plot the
orbits found.

In section II, we derive the geodesic equations of motion
using the variational problem associated with the metric of the
black hole space-time found in~\cite{AyonBeato:2009nh}. For both
massive particles and photons the following issues are investigated:
i) the effective potential of the equivalent one-dimensional
problem, which allows us to restrict the motions if they are radial
or presents a non-zero angular momentum, ii) the exact solution for
radial geodesics in terms of the proper and coordinate time, and
iii) exact solutions for the polar equation in motions with angular
momentum. The exact solutions are given in terms of the Weierstrass
elliptic $\wp$, $\sigma$, and $\zeta$ functions. In section III we
summarize and discuss our results.

\section{Geodesics of the Lifshitz black hole space-time}
The three-dimensional Lifshitz black hole solutions
presented in Ref. \cite{AyonBeato:2009nh} are a family (which generic
values of the dynamical exponent $z$) described by the following
metric:
\begin{equation}
ds^2=-\frac{r^{2z}}{\ell^{2z}}F(r)dt^2+\frac{\ell^{2}}{r^{2}}H(r)dr^2+r^{2}d\phi^2, \label{lif1}
\end{equation}
where the coordinate are defined as  $-\infty \leq t \leq \infty$,
$r\geq0$, $0\leq\phi\leq 2\pi$ and functions $F(r)$ and $H(r)$
depends only of the radial coordinate. Besides, these
function take the value $F(r)=H(r)=1$ when $r \rightarrow \infty $,
which means that the metrics are asymptotically Lifshitz. 
In order to have a single horizon it was required that these functions present a
single-zero at location of the horizon ($r_+$). For the case
$z=3$ the equations of motions are solved choosing the following form
for the functions $F(r)$ and $H(r)$ \cite{AyonBeato:2009nh}:
\begin{equation}
F(r)=H^{-1}(r)=\left(1-\frac{M\ell^2}{r^2}\right)
\end{equation}
where $M$ is an integration constant and $\ell$ is the length
associated to the three-dimensional cosmological constant. This
solution corresponds to the static asymptotically Lifshitz black
hole, whose metric can be written as
\begin{equation} ds^{2}=-\frac{r^{4}\Delta}{\ell^{6}}dt^{2}+\frac{\ell^{2}}{\Delta}dr^{2}+r^{2} d\phi^{2},
\label{e.1} \end{equation} where $\Delta=r^{2}-r_{+}^{2}$. The
Kretschmann scalar associated to this metric is given by
\begin{equation}\label{kresh}
R_{\mu\nu\lambda\sigma}R^{\mu\nu\lambda\sigma}=
-\frac{4\left(8r_{+}^{4}-48r_{+}^{2}r^{2}+91r^{4}\right)}{\ell^{4}r^{4}}
\end{equation}
which diverges for $r\rightarrow 0$. For $r \rightarrow
\infty $ the scalar takes the value $ -364/\ell^{4}\sim
\lambda^{2}$, which characterize an AdS space-time. The curvature
invariants $R$ and $R_{\mu\nu}R^{\mu\nu}$ also diverge at
$r=0$~\cite{AyonBeato:2009nh} indicating a curvature singularity at
the origin. A single horizon is located at $r_+=\ell\,\sqrt{M}$. The
case $z=1$ described the well known BTZ black hole~\cite{BTZ}. The
geodesic motion on this space-time was first studied in~\cite{Gamboa}
and a general study of its structure geodesic was explored
in~\cite{cmp}.

Let us mention some differences of the space-time described by the
metric (\ref{e.1}) and the BTZ black hole. Contrary to our intuition
in $3+1$ dimensions where gravity increases as the distance to the
center of force diminishes (and so does the curvature), which is a property
of the Schwarzschild metric; in the BTZ black hole the curvature is
constant and negative. Nevertheless, the Lifshitz black hole
solution has a variable curvature as Eq.(\ref{kresh}) shows. For the
BTZ black hole $M$ is the conserved charge associated with
asymptotic invariance under displacement, so it can be identified with
the black hole mass. In the case of Lifshitz black hole, its mass
has been evaluated in~\cite{Myung}, using the Euclidean action
approach~\cite{Gonzalez} with the action given in ~\cite{NMG}. The
expression found for the mass, $\mathcal{M}$, is
\begin{equation}\label{mass}
\mathcal{M}=\frac{r_+^{4}}{4G_{3}\ell^4}= \frac{M^{2}}{4G_{3}},
\end{equation}
indicating the relationship between the constant of
integration which appears in the metric (\ref{e.1}) and the black
hole mass.

In order to compute the geodesic structures of the
Lifshitz black hole, we start from the Lagrangian given by
\begin{equation}
2\mathfrak{L}=
-\frac{r^{4}\Delta}{\ell^{6}}\dot{t}^{2}+\frac{\ell^{2}}{\Delta}\dot{r}^{2}+r^{2}\dot{\phi}^{2}=
-h,\label{gs1}\end{equation} where a dot corresponds to the derivative
with respect to an affine parameter, $\tau$, along the trajectory,
and $h=0$ for massless particles and $h=1$ for massive
particles.
Considering that the Lagrangian (\ref{gs1}) is independent
of $(t, \phi)$, their corresponding conjugate
momenta, $(\Pi_t, \Pi_{\phi})$, are conserved, and
are given by
\begin{equation} \sqrt{E}=\frac{r^{4}\Delta}{\ell^{6}}\dot{t}, \qquad
\textrm{and} \qquad L=r^{2}\dot{\phi},
\label{gs2}\end{equation}
respectively. Using these constants of motion in the radial
equation we can write the radial quadratures in ($\tau, t, \phi$) as
\begin{equation} \left(\frac{dr}{d\tau}\right)^{2}=
\frac{\ell^{4}}{r^{4}}
\left[E-V_{Lif}(r)\right],\label{gs3}\end{equation}
\begin{equation} \left(\frac{dr}{dt}\right)^{2}=\frac{r^{4}\Delta^{2}}{\ell^{8}E^{2}}
\left[E-V_{Lif}(r)\right],\label{gs4}\end{equation}
and
\begin{equation} \left(\frac{dr}{d\phi}\right)^{2}=
\frac{\ell^{4}}{L^{2}}
\left[E-V_{Lif}(r)\right],
\label{gs5}\end{equation}
where $V_{Lif}(r)$ is the so-called effective potential for the 2+1 Lifshitz
space-time, which is given by
\begin{equation} V_{Lif}(r)=
\frac{r^{4}\Delta}{\ell^{6}}\left(h+\frac{L^{2}}{r^{2}}\right).
\label{gs6}\end{equation}

It is well known that an extreme of the effective potential, located
at some point $r_{c,h}$, gives us information on the existence and
stability of the circular orbits, and also information on the
structure of space-time. For example, under certain circumstances the
existence of a stable circular orbit would give the possibility to
have confined orbits. Thus, considering (\ref{gs6}) we obtain the
following equality:
\begin{equation}\label{gs7}
  L^2 \,(2 \,r_{c,h}^2 -  r_+^2) + h\, r_{c,h}^2 (3\, r_{c,h}^2 - 2   r_+^2)=0,
\end{equation}
which for massless particles has the solution
\begin{equation}
r_{c,0}=\frac{r_+}{\sqrt{2}}\equiv R_c\approx 0.71\,r_+.
\label{gs8}\end{equation} This result is independent of the angular
momentum and smaller than the event horizon. On the other hand, for
massive particles we obtain
\begin{equation}\label{gs9}
  r_{c,1}=\sqrt{\frac{ r_+^2 + \sqrt{L^4 + L^2\, r_+^2 + r_+^4}-L^2}{3}},
\end{equation}
which is always smaller than the event horizon, independent of the
value of the angular momentum. Also, note that
\begin{eqnarray}
  r_{c,1} &\rightarrow& R_c\,\left(1+\frac{1}{4}\,\frac{R_c^2}{L^2}+0(R_c/L)^3 \right)\approx 0.71\,r_+,\qquad \textrm{if}\qquad L\rightarrow \infty, \\
   r_{c,1} &\rightarrow&\frac{2\sqrt{3}}{3}\,R_c\,\left(1-\frac{1}{16}\,\frac{L^2}{R_c^2}+0(L/R_c)^3 \right)\approx 0.82\, r_+,\qquad \textrm{if}\qquad L\rightarrow 0.
\end{eqnarray}
Therefore, from (\ref{gs8}) and (\ref{gs9}) we conclude
that confined orbits are forbidden in the 2+1
Lifshitz background. This last fact has also been found recently in a
topological Lifshitz black hole in 3+1 dimensions \cite{germancito}.

\subsection{Time Like Geodesics }
\begin{figure}[!h]
 \begin{center}
  \includegraphics[width=85mm]{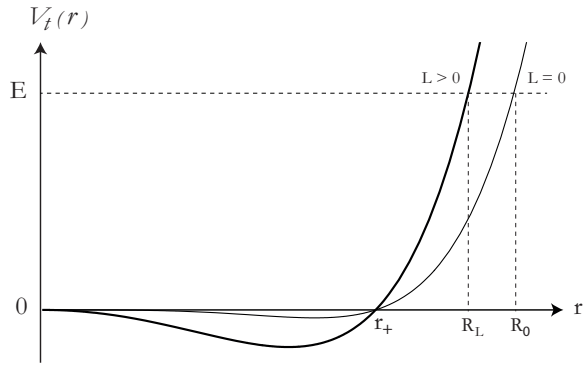}
 \end{center}
 \caption{Effective potential for massive particles. 
 As the angular momentum increase, the maximum distance 
 that the particle can reach diminishes, for a given $E$.}
 \label{Fig.1}
\end{figure}

In this case $h=1$, and the effective potential (\ref{gs6}) can be
rewritten in the following form:
\begin{equation} V_{t}(r)= \frac{r^{4}}{\ell^{6}}\left(r^{2}-r^{2}_{+}\right)
\left(1+\frac{L^{2}}{r^{2}}\right). \label{TL1}
\end{equation}
In Fig. \ref{Fig.1} we have plotted this potential for particles with
null angular momentum $L=0$ and non-vanished angular momentum $L\neq
0$.

\subsubsection{Angular Motion}
Angular motions of massive particles are characterized by $L>0$, such that
the effective potential yields
\begin{equation} V_{ta}(r)= \frac{r^{4}}{\ell^{6}}\left(r^{2}-r^{2}_{+}\right)
\left(1+\frac{L^{2}}{r^{2}}\right).\label{ta1}\end{equation}

Thus, the turning point, $R_{L}$, is given by

\begin{equation}R_L^{2}= \frac{2}{3}\sqrt{r_{+}^{4}+r_{+}^{2}L^{2}+L^{4}}
 \cosh\left[\frac{1}{3}\cosh^{-1}\left(\frac{27E\ell^{6}+2r_{+}^{6}+3r_{+}^{4}L^{2}-3r_{+}^{2}L^{4}-2L^{6}}
{2\sqrt{(r_{+}^{4}+r_{+}^{2}L^{2}+L^{4})^{3}}}\right)\right]
+\frac{r_{+}^{2}-L^{2}}{3},\label{ta2}\end{equation}
and the quadrature (\ref{gs3}) can be written as
\begin{equation} \phi(r)=L\ell \int_{R_L}^{r} \frac{-dr}{\sqrt{E\ell^{6}+r_{+}^{2}L^{2}r^{2}+(r_{+}^{2}
-L^{2})r^{4}-r^{6}}}.
\label{ta3}\end{equation}

\begin{figure}[!h]
 \begin{center}
  \includegraphics[width=55mm]{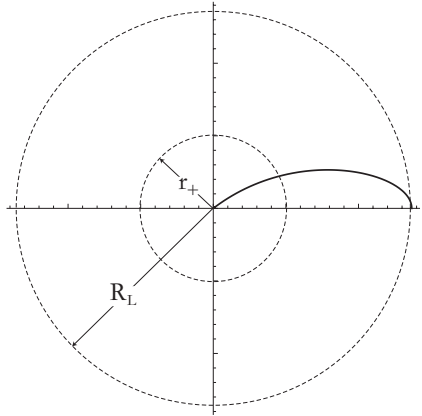}
 \end{center}
 \caption{Polar plot of the trajectories for massive particles with non-vanished angular momentum $L>0$.}
 \label{f2}
\end{figure}

Therefore, after an integration we obtain the polar
trajectory,
\begin{equation} r(\phi)=\frac{\ell}{
\sqrt{4\wp(k\phi+\varpi; g_{2},g_{3})-\delta_0}},
\label{ta4}\end{equation}
where $\wp(y; g_{2},g_{3})$ is the P-Weierstrass function,
$g_2$ and $g_3$ are the so-called Weierstrass invariants given by

\begin{equation} g_{2}=\frac{r_{+}^{4}L^{4}}{12 E^{2}\ell^{8}}-\frac{r_{+}^{2}-L^{2}}{4E\ell^{2}},\qquad \textrm{and}\qquad
g_{3}=\frac{1}{16E}+\frac{r_{+}^{2}L^{2}(r_{+}^{2}-L^{2})}{48E^{2}\ell^{6}}-\frac{r_{+}^{6}L^{6}}{216E^{3}\ell^{12}},
\label{ta5}\end{equation}
where the constants are
\begin{equation}
\delta_0 =\frac{r_{+}^{2} L^{2}}{3E\ell^{4}},
\qquad k=\frac{2\ell\sqrt{E}}{L},\qquad\textrm{and}\qquad
\varpi=\left|\wp^{-1}\left(\frac{1}{4}\left(\delta_0+\frac{\ell^{2}}{R_L^{2}}\right); g_{2},g_{3}\right)\right|.
\label{ta6}\end{equation}
The polar trajectory (\ref{ta4}) is plotted in Fig. \ref{f2}.

\subsubsection{Radial Motion}
In this cases we find that $L=0$ and the
effective potential (\ref{TL1}) becomes
\begin{equation}
V_{tr}(r)= \frac{r^{4}}{\ell^{6}}\left(r^{2}-r^{2}_{+}\right),
\label{tr1}\end{equation} whose behaviors is showed in Fig. \ref{Fig.1},
and the quadrature (\ref{gs3}) can be written as
\begin{equation}
\tau(r)=\ell \int_{R_{0}}^{r} \frac{-r^{2} dr}
{\sqrt{(R_{0}^{2}-r^2)[r^4+(R_{0}^{2}-r^{2}_{+})r^2+R_{0}^{2}(R_{0}^{2}-r^{2}_{+})]}}.
\label{tr2}\end{equation} The turning point, $R_{0}$, is given by
\begin{equation}
R_0 ^{2}= \frac{r_{+}^{2}}{3}+\frac{2 r_{+}^{2}}{3}
 \cosh\left[\frac{1}{3}\cosh^{-1}\left(1+\frac{27E\ell^{6}}{2\,r_{+}^{6}}
\right)\right].
\label{tr3}\end{equation}
So, after an integration of Eq. (\ref{tr2}) we obtain
\begin{equation}
\tau(r)=\frac{\ell R_{0}}{2\sqrt{3 R_{0}^{2}-2 r^{2}_{+}}}
\left[\left(1+\frac{\zeta(\Omega)}{2\wp^{'}(\Omega)}\right)\wp^{-1}(U)
+\frac{1}{4\wp^{'}(\Omega)}\ln\left(\frac{\sigma(\wp^{-1}(U)-\Omega)}
{\sigma(\wp^{-1}(U)+\Omega)}\right)
\right],\label{tr4}\end{equation}
where $\wp^{-1}(Y)\equiv\wp^{-1}(Y; g_2, g_3)$ is the inverse
P-Weierstrass function, $\sigma(Y)\equiv \sigma(Y; g_2, g_3)$ is the sigma Weierstrass function,
$\zeta(Y)\equiv \zeta(Y; g_2, g_3)$ is the zeta Weierstrass function,
and the Weierstrass invariants are given by
\begin{equation}
g_2={1\over12}(A^2-A+1-3B),\qquad
g_3={1\over432}(A-2)(2A^2+A-1-9B),
\label{tr5}\end{equation}
with
\begin{equation}
A={2 R_{0}^{2}-r^{2}_{+} \over 3 R_{0}^{2}-2 r^{2}_{+}},\qquad \textrm{and}\qquad
B={ R_{0}^{2}\over 3 R_{0}^{2}-2 r^{2}_{+}}.
\label{tr5}
\end{equation}
Also, the radial function $U=U(r)$ and the constant $\Omega$
are given by

\begin{equation}
U(r)={1 \over 4}\left[{R_{0}^{2} \over R_{0}^{2}-r^{2}}
-{2 R_{0}^{2}-r^{2}_{+} \over 3 R_{0}^{2}-2 r^{2}_{+}}\right],
\qquad \textrm{and}\qquad \Omega=\wp^{-1}\left({1+A \over 12}\right).
\label{tr6}
\end{equation}

On the other side, from Eq.(\ref{gs4}) we obtain the quadrature
\begin{equation}t(r)=\ell^{7}\,\sqrt{E} \int_{R_{0}}^{r} \frac{-dr}
{r^{2}(r^{2}-r^{2}_{+})\sqrt{E\ell^{6}+r_{+}^{2}r^{4}-r^{6}}}.
\label{tr7}\end{equation}
Thus, making the change of variable $ x = \ell^ {2} / r^{2} $,
and then integrating the remaining expression, we obtain
\begin{equation} t(r)=\frac{\ell^{5}}{r_{+}^{4}}
\left\{\frac{\wp(w)}{\wp^{'}(w)}\left[2\left(\alpha-\wp^{-1}\left(\frac{\ell^{2}}{r^{2}}\right)\right)\zeta(w)
+\beta-\ln\left(\frac{\sigma\left(\wp^{-1}\left(\frac{\ell^{2}}{r^{2}}\right)-w\right)}
{\sigma\left(\wp^{-1}\left(\frac{\ell^{2}}{r^{2}}\right)+w\right)}\right)\right]
+\alpha-\wp^{-1}\left(\frac{\ell^{2}}{r^{2}}\right)
+\frac{\zeta[\wp^{-1}\left(\frac{\ell^{2}}{r^{2}}\right)]-\zeta(\alpha)}{\wp(w)}\right\}.
\label{tr8}\end{equation}
In this case the Weierstrass invariants are given by
\begin{equation}
g_2=-\frac{4r_{+}^{2}}{E\,\ell^{2}},\qquad \textrm{and}\qquad
g_3=\frac{4}{E},
\label{tr9}\end{equation}
together with the constants
\begin{equation}
w=\wp^{-1}\left(\frac{\ell^{2}}{r_{+}^{2}}\right),\qquad \alpha=\wp^{-1}\left(\frac{\ell^{2}}{R^{2}_{0}}\right),
\qquad \textrm{and}\qquad \beta=\ln\left(\frac{\sigma(\alpha-w)}{\sigma(\alpha+w)}\right).
\label{tr10}
\end{equation}
\begin{figure}[!h]
 \begin{center}
  \includegraphics[width=85mm]{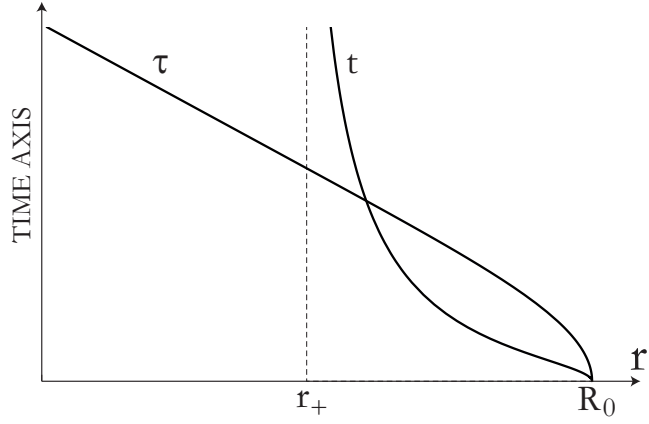}
 \end{center}
 \caption{Radial massive particles in the proper and coordinate time.}
 \label{f3}
\end{figure}
In Fig. \ref{f3} we plot the proper time, Eq. (\ref{tr4}), and the
coordinate time, Eq. (\ref{tr8}), as a function of the radial coordinate, $r$.
We can see that  radial massive particles cross the event horizon
in a finite proper time. However, an external observer will
see that the particle falls asymptotically toward the event horizon.
This result is common with the $3+1$ Einstein space-times (Schwarzschild \cite{chandra},
SAdS \cite{COV}, etc.)


\subsection{Null Geodesics}
\begin{figure}[!h]
 \begin{center}
  \includegraphics[width=85mm]{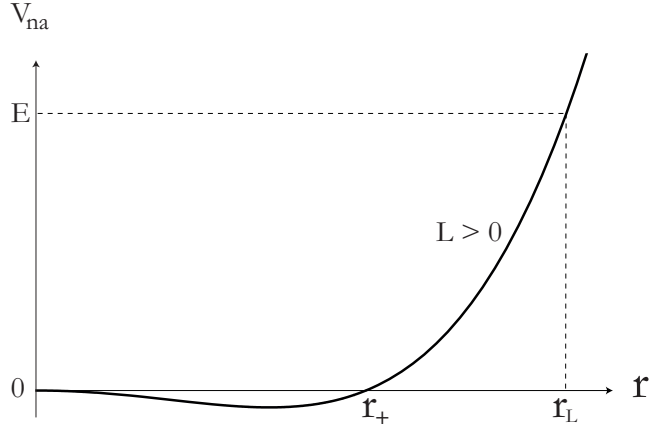}
 \end{center}
 \caption{Effective potential for massless particles with non-vanished angular momentum $L>0$.}
 \label{f4}
\end{figure}

The motion of massless particles is described by the effective
potential (\ref{gs6}) with $h=0$,
\begin{equation}
V_{n}(r)= \frac{L^{2}}{\ell^{6}}\left(r^{2}-r^{2}_{+}\right)r^{2},
\label{n1}\end{equation} which we draw in Fig. \ref{f4}.  As we have
shown, confined orbits are forbidden in this space-time, even though
we can determine explicitly the trajectories of massless particles.

\subsubsection{Angular Motion}
\begin{figure}[!h]
 \begin{center}
  \includegraphics[width=60mm]{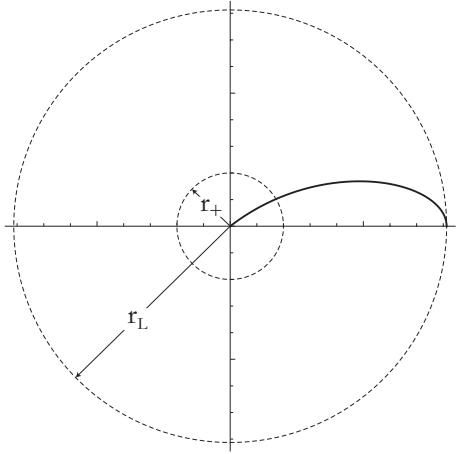}
 \end{center}
 \caption{Polar plot for massless particles with $L>0$.}
 \label{f5}
\end{figure}

From Eq. (\ref{gs5}), with $L>0$ and defining the impact parameter
as $b=L/E$, it is possible to write the angular quadrature as
\begin{equation}
\phi(r)=-\int_{r_{L}}^{r} \frac{dr}
{\sqrt{\frac{\ell^{4}}{b^{2}}-\frac{r^{2}}{\ell^{2}}(r^{2}-r^{2}_{+})}}
= - \ell \int_{r_{L}}^{r}
\frac{dr}{\sqrt{(r_{L}^{2}-r^{2})(\rho_{L}^{2}+r^{2})}},
\label{n2}\end{equation}
where $r_L$ is the turning point, and
$\rho_L$ is a negative root without physical meaning, which are
given by
\begin{equation} r_{L}=r_{+}
\left[\frac{1}{2}+\sqrt{\frac{1}{4}+\frac{\ell^{6}}{r_{+}^{4}b^{2}}}\right]^{1/2}\qquad
\rho_{L}=r_{+}
\left[-\frac{1}{2}+\sqrt{\frac{1}{4}+\frac{\ell^{6}}{r_{+}^{4}b^{2}}}\right]^{1/2},
\label{n3}\end{equation}
respectively.
Thus, integrating Eq. (\ref{n2}) the polar form of the trajectory is obtained
in terms of the P-Weierstrass function
\begin{equation}
r(\phi)= r_{L}-\frac{\ell}{4\wp(\kappa\, \phi; g_{2},
g_{3})+\delta}, \label{n4}\end{equation}
where the Weierstrass
invariants are given in this case by
\begin{equation}
g_{2}=-\frac{\ell^{2}(14\,r_{L}^{2}\,\rho_{L}^{2}-r_{L}^{4}-\rho_{L}^{4})}{48\,r_{L}^{2}(r_{L}^{2}+\rho_{L}^{2})^{2}},
\qquad
g_{3}=-\frac{\ell^{3}(r_{L}^{2}-\rho_{L}^{2})(r_{L}^{4}+34\,r_{L}^{2}\,\rho_{L}^{2}+\rho_{L}^{4})}{1728\,r_{L}^{3}(r_{L}^{2}+\rho_{L}^{2})^{3}},
\label{n5}\end{equation}
where the constants are
\begin{equation}
\kappa=\sqrt{\frac{2r_{L}(r_{L}^{2}+\rho_{L}^{2})}{\ell^{3}}}\qquad
\delta= \frac{\ell}{6\,r_{L}}+\frac{2\ell
r_{L}}{3(r_{L}^{2}+\rho_{L}^{2})}. \label{n6}\end{equation} In
Fig.\ref{f5} we plot the photons trajectory given by expression
(\ref{n4}).

\subsubsection{Radial Motion}
\begin{figure}[!h]
 \begin{center}
  \includegraphics[width=85mm]{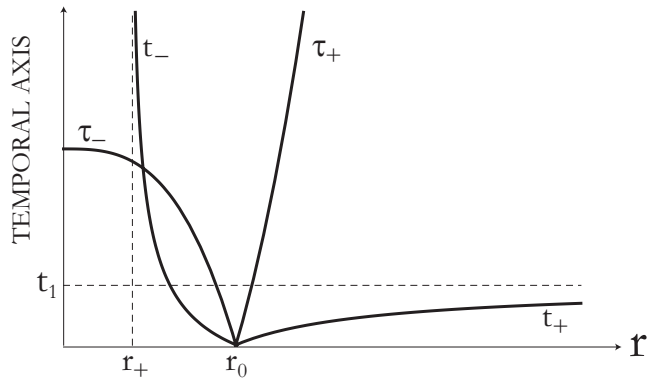}
 \end{center}
 \caption{Radial massless particles in the proper and coordinate time.}
 \label{f6}
\end{figure}

In the case of the radial motion of photons ($L=0$), the effective
potential (\ref{n1}) becomes
\begin{equation}\label{n7}
  V_{nr}=0,
\end{equation}
and therefore, radial photons can escape to
infinity. The dynamic equations are given by
\begin{equation}
\pm\frac{dr}{d\tau}=\frac{\ell^{2}\sqrt{E}}{r^{2}},
\label{n8}\end{equation}
and
\begin{equation}
\pm\frac{dr}{dt}=\frac{r^{2}\Delta}{\ell^{4}}.
\label{n9}\end{equation}
Integrating Eq.(\ref{n8}) we obtain
\begin{equation}
\tau(r)=\pm\frac{1}{3\,\ell^2 \sqrt{E}}(r^{3}-r_{0}^{3}).
\label{n10}\end{equation} The minus sign represents photons that are
falling toward the black hole, taking a finite proper time,
$\tau_+\equiv \tau(r_+)$, to cross the event horizon; the plus sign
means that the photons escape to spatial infinity, taking an
infinite proper time (see Fig. \ref{f6}).

On the other hand, the quadrature (\ref{n9}) can be written as
\begin{equation}
t(r)=\pm\ell^{4}\int_{r_{0}} ^{r}
\frac{dr}{r^{2}(r^{2}-r_{+}^{2})},
\label{n11}\end{equation}
so, after a brief manipulation in partial fractions,
and then integrating, we find that the solution is
\begin{equation}
t(r)=\pm\ell^4\left[\frac{1}{2r_{+}^{3}}\ln\left|\frac{(r-r_{+})(r_{0}+r_{+})}{(r+r_{+})(r_0-r_{+})}\right|
+\frac{1}{r_{+}^{2}}\left(\frac{1}{r}-\frac{1}{r_{0}}\right)
\right]. \label{n12}\end{equation} This last expression shows that
photons traveling to $r_+$ have a similar behavior as the static
spherically  symmetric space-times of Einstein's theory, i. e.,
an external system will see that photons fall asymptotically to
$r_+$. Moreover, as we have mentioned above, photons take an infinite
proper time traveling to infinity, but an external system will see
that photons arrives at infinite in a finite coordinate time, $t_1$,
which is given by
\begin{equation}\label{n13}
  t_1=\lim \limits_{r\rightarrow \infty} t(r)=\ell^4\left[\frac{1}{2r_{+}^{3}}\ln\left(\frac{r_{0}+r_{+}}{r_0-r_{+}}\right)
  -\frac{1}{r_{+}^{2}\,r_{0}}\right].
\end{equation}
This behavior is shown in Fig. \ref{f6}, and has been
reported recently in the topological Lifshitz 3+1 counterpart
\cite{germancito}.

\section{Final Remarks}

In this work we have studied the geodesic structure
of the Lifshitz black hole in $2+1$ dimensions by means of an analysis of the motion
of massive and massless particles in this background. Using the standard
Lagrangian procedure we obtain the equations of motion in terms of the
energy, $E$, and angular momentum, $L$, which as a result are found to be constants of motion
for the particles. Also, we obtain an effective potential, $V(r)$, that allows us
to describe qualitatively the movement, and then, we obtain
an analytical solution to equations of motion.

In the first place, we found that confined orbits are forbidden (except
for radial motion of massless particles), because the minimum of the
effective potential as a result is found to be smaller than the event horizon,
$r_+$.

For massive particles we found that the radial motion is similar
to the Schwarzschild case: in the proper system, they cross
the horizon in a finite time, whereas an external system will
observe that they take an infinite time to do it. We also found the
trajectories of massive particles falling into the horizon with
$L>0$ in terms of the elliptic $\wp$-Weierstrass function.

On the other side, the motion of massless particles with $L>0$ has
the same general behavior of massive particles. Moreover, photons
with vanishing angular momentum can escape to spatial infinity: in
the proper system they take an infinity of time to escape, but an 
external system will observe that they take a finite time, $t_1$,
to do it, as shown in Eq. (\ref{n13}). This last fact also has been
found in the Topological Lifshitz black hole \cite{germancito}.

\begin{acknowledgments}
M. O. thanks to PUCV.
This work was supported by DICYT-USACH Grant No 041331CM (NC).
\end{acknowledgments}



\begin{thebibliography}{9}

\bibitem{AyonBeato:2009nh}
Ayon-Beato E., Garbarz A., Giribet G. and Hassaine M.:
Lifshitz Black Hole in Three Dimensions.
 Phys. Rev.  {\bf D} 80, 104029 (2009) [arXiv: 0909.1347].

\bibitem{Balasu}
Balasubramanian K. and McGreevy J.:
An analytic Lifshitz black hole.
Phys. Rev. {\bf D} 80, 104039 (2009) [arXiv: 0909.0263].

\bibitem{BTZ}
Ba\~{n}ados M., Teitelboim C. and Zanelli J.: The Black
Hole in three-dimensional Space Time.
Phys. Rev. Lett. \textbf{69}, 1849 (1992) [arXiv: 9204099].;
Ba\~{n}ados M., Henneaux M., Teitelboim C. and Zanelli J.:
Geometry of the 2+1 Black Hole.
Phys. Rev. \textbf{D} 48, 1506 (1993) [arXiv: 9302012].

\bibitem{NMG}
Bergshoeff E., Hohm O. and Townsend P.:
Massive Gravity in Three Dimensions.
Phys. Rev. Lett. \textbf{102} 201301 (2009) [arXiv: 0901.1766].

\bibitem{DanielssonII}
Brynjolfsson E., Danielsson U., Thorlacius L. and Zingg T.:
Holographic Superconductors with Lifshitz Scaling.
J. Phys. {\bf A} 43, 065401 (2010) [arXiv: 0908.2611].

\bibitem{chandra}
Chandrasekhar S.:
The Mathematical Theory of Black Holes.
Oxford University Press, New York (1983).

\bibitem{cmp}
Cruz N., Mart\'inez C. and Pe\~na L.:
Geodesic structure of the (2 + 1)-dimensional BTZ black hole.
Class. Quantum Grav. \textbf{11}, 2731 (1994) [arXiv: 9401025].

\bibitem{COV}
Cruz N., Olivares M. and Villanueva J. R.:
The geodesic structure of the Schwarzschild anti-de Sitter Black Hole.
Class. Quantum Grav. \textbf{22}, 1167-1190 (2005) [arXiv: 0408016].


\bibitem{DanielssonI}
Danielsson U. and Thorlacius L.:
Black holes in asymptotically Lifshitz space-time.
JHEP {\bf 0903}, 070 (2009) [arXiv: 0812.5088].

\bibitem{Gamboa}
Farina C., Gamboa J. and Segu\'{\i}-Santonja A. J.:
Motion and Trajectories of Particles Around Three-Dimensional Black Holes.
Class. Quantum Grav. \textbf{10}, 193 (1993) [arXiv: 9303005].

\bibitem{Gonzalez}
Gonzalez H. A., Tempo D. and Troncoso R.:
Field theories with anisotropic scaling in 2D, solitons and the microscopic entropy of asymptotically Lifshitz black holes.
\textit{JHEP}, \textbf{1111}, 066 (2011) [arXiv: 1107.3647].


\bibitem{Kachru}
Kachru S., Liu X. and Mulligan M.:
Gravity Duals of Lifshitz-like Fixed Points.
Phys. Rev. \textbf{D} 78, 106005 (2008) [arXiv: 0808.1725].


\bibitem{Mann}
Mann R. B.:
Lifshitz topological black holes.
JHEP \textbf{06}, 075 (2009). [arXiv: 0905.1136].

\bibitem{Myung}
Myung Y. S.:
Phase transitions for the Lifshitz black holes.
Eur. Phys. J. {\bf C} 72 2116 (2012) [arXiv: 1203.1367].

\bibitem{Nakasone}
Nakasone M. and Oda I.:
On Unitarity of Massive Gravity in Three Dimensions.
Prog. Theor. Phys. {\bf 121}, 1389 (2009) [arXiv: 0902.3531];
Deser S.:
Ghost-free, finite, fourth order D=3 (alas) gravity.
Phys. Rev. Lett. {\bf 103}, 101302 (2009) [arXiv: 0904.4473].

\bibitem{germancito}
Olivares M., Rojas G., V\'asquez Y. and Villanueva J. R.:
Particles motion on topological Lifshitz black holes in 3+1 dimensions.
Accepted for publication in Astrophys. Space Sci. (2013) [arXiv: 1304.4297].




\end{thebibliography}
\end{document}